\documentstyle[psfig]{lamuphys}
\makeatletter
\let\chapter\hid@chapter
\makeatother

\begin{document}
\pagenumbering{arabic}

\title{Study of the LISM using Pulsar Scintillation}

\author{
N. D. Ramesh Bhat,  
Yashwant Gupta, \and
A. Pramesh Rao
}

\institute{
National Centre for Radio Astrophysics (TIFR), Pune 411 007, India.
}

\maketitle

%
%
%

\begin{abstract}

We present here the results from an extensive scintillation study of twenty pulsars 
in the dispersion measure (DM) range $ 3-35 \ {\rm pc \ cm^{-3}}$ carried out using 
the Ooty Radio Telescope, to investigate the distribution of ionized material in the 
local interstellar medium (LISM). 
Our analysis reveals several anomalies in the scattering strength, which suggest that 
the distribution of scattering material in the solar neighborhood is not uniform.
Our model suggests the presence of a low density bubble surrounded by a shell of much 
higher density fluctuations. 
We are able to put some constraints on geometrical and scattering properties of such a 
structure, and find it to be morphologically similar to the local bubble known from 
other studies.

\end{abstract}

\section{Introduction}

Propagation effects on pulsar signals, such as dispersion and scattering, 
probe the distribution of thermal plasma in the interstellar medium (ISM). 
Scattering studies of radio waves from pulsars, enable us to probe the 
electron density fluctuations in the ISM, which are presumably due to 
turbulence in the ISM (Rickett 1990).
Not much is known about electron densities and their fluctuations in the 
LISM and in order to investigate this in detail, we have 
made extensive scintillation observations of twenty nearby pulsars.
Reliable and accurate estimates of strength of scattering have been obtained
and the results are used to study the distribution of electron density 
fluctuations in the LISM.

\section{Observations and Data Analysis}

The observations were made using the Ooty Radio Telescope (ORT) at 327 MHz. 
Within the sky coverage and sensitivity limits of the ORT, there are twenty 
nearby pulsars that were found suitable for studying the LISM.
For these pulsars, the dynamic scintillation spectra $-$ intensity variation 
in the frequency-time plane $-$ were measured over 20$-$60 epochs, spanning 
$ \sim $ 100 days during 1993$-$95.
Such spectra display intensity scintillation patterns that fade over narrow 
frequency ranges and short time intervals (Fig. 1), arising from diffractive
scintillation effects.
We obtained dynamic spectra data with a frequency resolution $ \approx $ 140 
kHz and a time resolution $ \sim $ 10 secs. 

\begin{figure}[thbp]
{\psfig{file=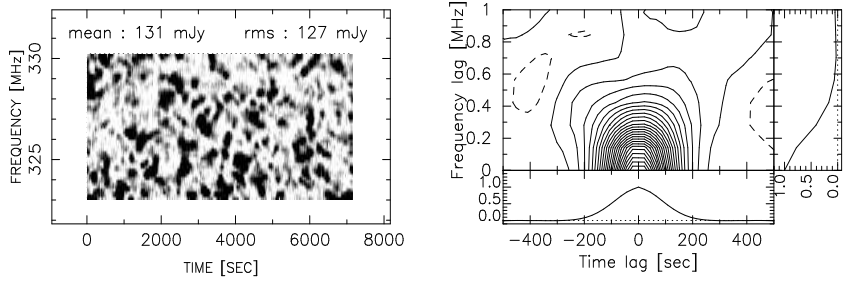,rheight=3.5cm,rwidth=5.0cm,height=9.5cm,width=12.0cm,angle=270}}
\caption[]{The dynamic scintillation spectrum of PSR B0919+06 as observed on 
           24 May 1994, darker areas representing higher intensity values 
	   (left panel). The ACF is shown along with 1D cuts 
	   across zero lags of frequency and time (right panel).}
\label{c33_fig01_fin}
\end{figure}

To quantify the average characteristics of scintillation patterns at any epoch,
we have computed the two-dimensional auto co-variance function (ACF) and fitted a 
two-dimensional gaussian to yield decorrelation bandwidth $ \nu _d $ and scintillation 
time scale $ \tau _d $, which are the widths of the ACF along zero time and frequency 
lag axes respectively (Gupta {\it et al}. 1994).
The large number of epochs of observations allow us to average out long-term 
fluctuations of these parameters, arising from refractive scintillation effects, 
giving an accuracy of 5$-$10\%. 
From these estimates, we derive the line-of-sight averaged strength of scattering 
$ \overline { C_n^2 } $ (Cordes {\it et al}. 1985), given by
\begin{equation}
\overline { C_n^2 } ~ \propto ~ \nu  _{obs} ^{\alpha } ~ D ^{- \alpha / 2}
 ~ \nu _d ^{ - ( { \alpha - 2 \over 2 } ) }
\end{equation}
where $ \nu _{obs} $ is the observing frequency, D is the distance estimate of the pulsar
(based on the model given by Taylor and Cordes 1993),
and $ \alpha $ is the power-law index for electron density spectrum (Armstrong {\it et al}. 1995).
We have assumed $ \alpha ~ = ~ { 11 \over 3 } $ in our calculations.

The derived values of $ \overline { C_n^2 } $ show roughly two orders of magnitude fluctuations 
(ranging from $ 10 ^ {-4.8 \pm 0.02} $ to $ 10 ^ {-3.1 \pm 0.02} $ $ {\rm m ^ { -20 / 3 } } $), which 
is about 10 times larger than earlier estimates and much larger than that predicted from current 
models for the $ \overline { C_n^2 } $ distribution in the galaxy (Cordes {\it et al}. 1985).
Fig. 2.c shows these variations have a systematic trend, in that they are dominant for nearby 
(D $ \la $ 1.5 kpc) pulsars.
In addition to this, we find several cases where pulsars at comparable DMs or distances 
show remarkably different scintillation characteristics. 
In order to study such effects, we consider two pulsars at similar DMs, having decorrelation 
bandwidths $ \nu _{d_1} $ and $ \nu _{d_2} $, and define an anomaly parameter ($ A _{dm} $) as
\begin{equation}
A_{dm} ~ = ~ { ( \nu _{d_1} / \nu _{d_2} ) _{observed} \over ( \nu _{d_1} / \nu _{d_2} ) _{expected} }
\hspace{0.5in} ({\rm if} ~ A_{dm} < 1, ~ {\rm then} ~ A_{dm} ~ = ~ { 1 \over A_{dm} })
\end{equation}
A plot of the anomaly parameter against DM (Fig. 2.a) shows a systematic 
variation, implying density fluctuations are distributed in a structure
that is local and is asymmetrically located relative to the sun. 
The anomaly parameter as a function of the distance (Fig. 2.b) shows a similar 
behaviour.

\begin{figure}[thbp]
{\psfig{file=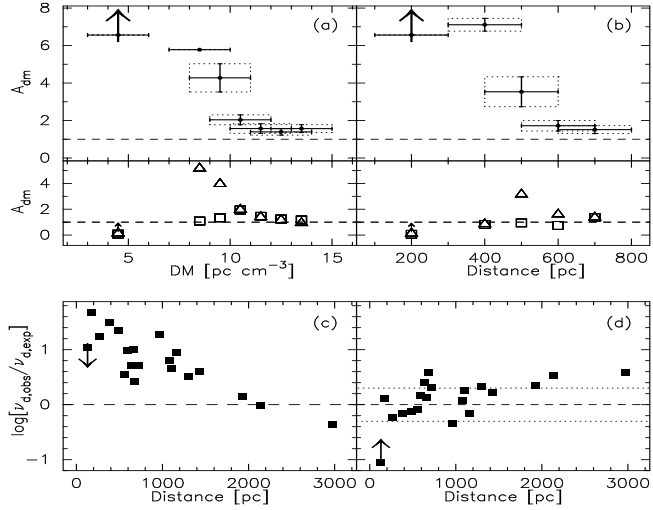,rheight=6.50cm,rwidth=7.0cm,height=7.5cm,width=13.0cm,angle=270}}
\caption[]{The variations of the parameter $ {\rm A _{dm} } $ are shown with DM (a) 
and with distance (b), where the expected values of $ \nu _d $ are for a uniform scattering 
medium. 
The lower panels of (a) and (b) are for the 3-component medium ($\triangle $ for uniform shell 
and $ \Box $ for non-uniform shell).
In (c), values of $ \nu _{d,exp} $ are computed for a uniform medium and in (d), they 
are for the 3-component medium (non-uniform shell).}
\label{c33_fig02_fin}
\end{figure}

\section{Interpretation of Results}

We try to understand our observations by comparing the measured anomalies
with anomalies predicted by specific density structures around the sun. 
Simple models consisting of an excess or deficiency of scattering material around the sun 
are unable to reproduce the observed trends of $ {\rm A_{dm}}$ and $ \nu _{d,obs} / \nu _{d,exp} $.
To explain our observations, we need a 3-component scattering medium with a cavity surrounded 
by a shell of much higher density fluctuations embedded in the normal, large scale ISM.
The sun has to be located away from the centre of the cavity.
By adjusting the parameters of the model, we are able to generate a reasonable agreement 
with the observations.
By further assuming that the strength of scattering from the shell decreases with z-height, 
we are able to improve the agreement (lower panels of Figs. 2.a and 2.b).
The best fit values of the parameters of the local scattering structure are such that 
the observed trends of the parameters $ {\rm A_{dm} } $ and $ \nu _d $ (Fig. 2) are reproduced.
The geometry of the structure is schematically shown in Fig. 3.
For the size,
constraints from our model are $a$ $\sim $ 55 pc, $b$ $\sim $ 60 pc
and $250 < c < 300$ pc. 
The centre has an offset $ r_c \sim $ 40 pc from the sun, towards $210^o < l < 235^o $ 
and $ -21^o < b < 21^o $.
The density fluctuations are such that 
$ 10^{-5} < \overline {C_n^2} < 10^{-4.45} ~ {\rm m^{-20/3} } $ for the inner cavity,
$ 10^{-1} < \int _0 ^d  C_n^2(l) ~ dl < 10^{-0.57} ~ {\rm pc ~ m^{-20/3} } $ for the shell, 
(where d is the thickness)
and $ \overline {C_n^2} < 10^{-3.37} ~ {\rm m^{-20/3} } $ for the outer ISM.
Since the strength of scattering of the shell is much higher than those of the cavity and the outer
ISM, it contributes substantially to the scattering of nearby pulsars outside the shell.
The relative contribution from the shell decreases with the distance, which gives rise to  
systematic variations of $ {\rm A _{dm} } $ and $ \nu _d $ (Fig. 2).
The offset from the sun ($ r_c $) and the size in the galactic plane ($a$ and $b$) are not 
uniquely constrained and our results can also be explained by a relatively bigger, but less 
asymmetric structure (the dashed geometry in Fig. 3). 

\begin{figure}[thbp]
{\psfig{file=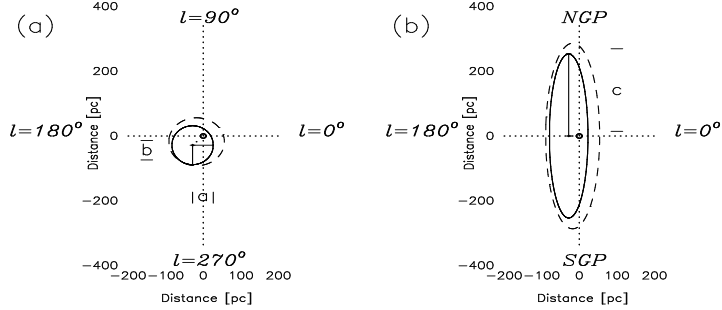,rheight=3.75cm,rwidth=5.0cm,height=6.00cm,width=10.0cm,angle=270}}
\caption[]{The model for the local scattering structure shown as
 	   cuts through the galactic plane (a) and NGP$-$SGP plane (b).
 	   The solid geometry is for $ r_c \sim $ 40 pc, $a \sim $ 55 pc and $b \sim $ 60 pc,
 	   and the dashed one has $ r_c \sim $ 25 pc and $ a \approx b \sim $ 75 pc.}
\label{c33_fig03_fin}
\end{figure}
  
The exact relations between the scattering strengths and other properties of the medium
are not well understood. 
If we assume $ C_n^2 \propto n_e^2 $, where $ n_e $ is the mean electron density, 
our observations would imply a density contrast $ \sim $ 10$-$20 times between 
the shell and the ambient ISM in the case of a thin shell ($ \sim $ 1 pc) and $ \sim $ 5$-$8 
times for $ \sim $ 10 pc thick shell. 
The simple model considered here does not constrain the thickness of the shell. 
The pulsar PSR B0950+08 that has a parallax distance $ \approx $ 130 pc is located, in our model,
within the cavity implying
$ n_e \approx 0.02 ~ {\rm cm^{-3}} $, which is 4 times larger than
its value from X-ray data (Snowden {\it et al}. 1990).
Our constraints are, morphologically, in broad agreement with that of local bubble,
obtained from X-ray, UV and HI data.
Our results suggest that the bubble is surrounded by a shell of much higher density fluctuations
and further investigations are required towards accurate 
estimates on more detailed characteristics of the shell boundary.


\begin{thebibliography}{}

\bibitem{}{ars95}{Armstrong et al.\ 1995}
Armstrong, J.W., Rickett, B.J., Spangler, S.R. (1995):
ApJ 443, 209

\bibitem{}{cwb85}{Cordes et al.\ 1985}
Cordes, J.M., Weisberg, J.M., Boriakoff, V. (1985):
ApJ 288, 221 

\bibitem{}{grl94}{Gupta et al.\ (1994)}
Gupta, Y., Rickett, B.J., Lyne, A.G. (1994):
MNRAS 269, 1035

\bibitem{}{r90}{}
Rickett B.J. (1990):
Ann. Rev. Astron. Ap. 28, 561

\bibitem{}{scms90}{Snowden et al.\ 1990}
Snowden, S.L., Cox, D.P., McCammon, D., Sanders, W.T. (1990):
ApJ 354, 211

\bibitem{}{tc93}{}
Taylor, J.H., Cordes, J.M. (1993):
ApJ 411, 674

\end{thebibliography}
\end{document}